\documentclass[aps,prd,preprint,superscriptaddress,nofootinbib,showpacs]{revtex4}
\usepackage{graphicx}

\usepackage{ulem}
\usepackage{color}
\definecolor{My_red}        {cmyk}{0.00,1.00,1.00,0.20}

\newcommand{\bmat}{\left(\begin{array}}
\newcommand{\emat}{\end{array}\right)}
\newcommand{\beq}{\begin{equation}}
\newcommand{\eeq}{\end{equation}}

\def\ra{\rightarrow}

\def\ld{\lambda}
\def\f{\frac}

\def\bwt{\begin{widetext}}
\def\ewt{\end{widetext}}
\def\be{\begin{equation}}
\def\ee{\end{equation}}
\def\bary{\begin{array}}
\def\eary{\end{array}}
\def\bit{\begin{itemize}}
\def\eit{\end{itemize}}

\def\ra{\rightarrow}

\def\ld{\lambda}

\def\su5u1{SU(5) \times U(1)}
\def\fsu5u1{SU(5) \times U(1)'}
\def\so10{SO(10)}
\def\sq20{SO(10) \times SO(10)}

\def\ra{\rightarrow}

\def\ld{\lambda}
\def\f{\frac}

\def\L{\left(}
\def\R{\right)}
\def\bwt{\begin{widetext}}
\def\ewt{\end{widetext}}
\def\be{\begin{equation}}
\def\ee{\end{equation}}
\def\bary{\begin{array}}
\def\eary{\end{array}}
\def\bit{\begin{itemize}}
\def\eit{\end{itemize}}

\def\ra{\rightarrow}

\def\ld{\lambda}

\def\su5u1{SU(5) \times U(1)}
\def\fsu5u1{SU(5) \times U(1)'}
\def\so10{SO(10)}
\def\sq20{SO(10) \times SO(10)}

\usepackage[centertags]{amsmath}
\usepackage{amssymb}

\begin{document}

\title{View FImP Miracle (by Scale Invariance) $\acute{\rm a}$ $\rm la$
 Self-interaction}

\author{Zhaofeng Kang}

\affiliation{ School of Physics, Korea Institute for Advanced Study,
Seoul 130-722, Korea}

\date{\today}

\begin{abstract}

Combining feebly interacting massive particle (FIMP) dark matter (DM) with scale invariance (SI) leads to extremely light FIMP (thus the FImP) with FImP miracle, i.e., the mass and relic generations of FImP DM share the same dynamics. In this paper we show that due to the lightness of FImP, it, especially for a scalar FImP, can easily accommodate large DM self-interaction. For a fermionic FImP, such as the sterile neutrino, self-interaction additionally requires a mediator which is another FImP, a scalar boson with mass either much lighter or heavier than the FImP DM. DM self-interaction opens a new window to observe FImP (miracle), which does not leave traces in the conventional DM searches. As an example, FImP can account for the offsets between the centroid of DM halo and stars of galaxies recently observed in the galaxy cluster Abel 3827.

\end{abstract}

\maketitle


The conventional weakly interacting massive particle (WIMP) paradigm for dark matter (DM) is being challenged by quite a few null results of direct and indirect DM searches. They are already probing the typical WIMP DM, and yield particularly strong bounds in the lighter DM region. On the other hand, viewing from DM relic density along, WIMP does not take advantage over FIMP, i.e., the feebly interacting massive particle~\cite{freezein0,freezein1}. Instead of the freeze-out dynamics of WIMP that is used to keep thermal equilibrium with the plasma, FIMP gains correct relic density $\Omega h^2\simeq0.1$ through the freeze-in dynamics. It never enters the plasma but still arrives $\Omega h^2\simeq0.1$ via slow thermal productions like thermal particles decay. An obvious merit of FIMP is that it, just as expected, leaves null results in the conventional DM detectors devised for WIMP DM. Moreover, as a competitor to the WIMP miracle, a miracle of FIMP arises from combining FIMP with scale invariance (SI)~\cite{Kang:2014cia}. This classical symmetry may provide a way to address the hierarchy problem~\cite{Bardeen}. 

FIMP with SI is a nontrivial combination, which gives rise to several important consequences. Immediately, quite generically FIMP must be extremely light (thus dubbed FImP). The point is simple. In most of the SI schemes for generating the electroweak (EW) scale, spontaneously breaking of SI happens at the electroweak (EW) scale~\cite{SI:boson} or TeV scale~\cite{SI:hidden}, by means of a scalar field collectively denoted as $\Phi$ with vacuum  expectation value (VEV) $u\equiv\langle \Phi\rangle \lesssim $ TeV. By virtue of SI, all particles including DM $X$ should gain masses via coupling to scalar fields with non-vanishing VEV, for instance, to $\Phi$. Schematically, we can write down terms for mass generation
\begin{align}\label{Laga}
\f{1}{2}\ld_\phi X^2\Phi^2,\quad \f{1}{2}y_\phi X^2 \Phi,
\end{align}  
with $X$ assumed to be a real scalar and Majorana fermion, respectively. Because $X$ is a FIMP,  one has $\ld_{\phi}, y_\phi\ll1$ thus very light FIMP. The ensuing important consequence is the aforementioned FImP miracle, which now becomes obvious: (due to SI) the mass and relic generations of FIMP share the common dynamics. The final consequence is that dark parity by hand may be not necessary for FImP. It can be sufficiently long-lived even without an exact protective symmetry, because its decay width is greatly suppressed by light mass and moreover feeble couplings.

FIMP barely produces observable signatures except for a decaying one~\cite{Kang:2010ha} (but may leave hint in tensor-to-scalar ratio~\cite{Dev:2014tla}). However, FImP does. Firstly, Since FImP, such as a sterile neutrino or Majoron~\cite{Queiroz:2014yna}, does not require a parity, it is well expected that it can decay into $X-$ray photon(s)~\footnote{Neutrino and photon are almost massless particles in the SM, so probably they are the only kinematically accessible final states for FImP decay. Neutrino is hard to observe while $X-$ray line is a good observable.}. Secondly, also the core of this paper, FImP is likely to have appreciable self-interaction~\footnote{DM self-interaction was originally motivated to address the small scale problems~\cite{Spergel:1999mh}. } partially by virtue of its lightness; hence, that kind of FImP can be probed from its astrophysical effects, e.g., leading to a separation between the DM halo and stars of a galaxy which is moving through a region with large DM number density. Interestingly, such a phenomenon was reported recently~\cite{Abel1}. It was discovered in the galaxy cluster Abell 3827, within which (in the inner 10 kpc core) four elliptical galaxies were observed and their DM halos were reconstructed, with at least one spatially offset from its stars by a distance of $\Delta = 1.62^{+0.47}_{-0.49}$ kpc~\cite{Abel1}. Such kind of offset can be explained by DM with self-interaction that leads to DM self-scattering rate per mass $\sigma_{\rm DM}/m_{\rm DM}\sim (1.0-1.5) \rm cm^2/g$~\cite{Abel2}, or in the particle physics unit:
\begin{align}\label{SI:cross}
\sigma_{\rm DM}/m_{\rm DM}\sim (4.7-7.0)  \times10^3\rm GeV^{-3}.
\end{align}  
Noticeably, despite of some tension with the bullet galaxy cluster upper bound, this value is also indicated to solve the small scale problems~\cite{Tulin:2013teo}. Therefore, it is of great interest to explore FImP with self-interactions (see other attempts~\cite{Campbell:2015fra,Heikinheimo:2015kra,Choi:2015bya}).

In the absence of velocity dependence, Eq.~(\ref{SI:cross}) is saying that the mass scale involved in DM scattering, mass of a mediator or DM itself, should be far below the weak scale. Actually, for a typical WIMP, ${\sigma_{\rm DM}}/{m}$ is expected to scale as 
\begin{align}\label{}
\f{\sigma_{\rm DM}}{m}\sim \f{1}{32\pi} \f{\ld^2}{m^3} \simeq 10^{-10}\times\L\f{\ld}{0.1}\R^2\L\f{100\rm GeV}{m}\R^3\rm GeV^{-3},
\end{align}  
which is about 14 orders of magnitude smaller than the indicated value given by Eq.~(\ref{SI:cross}). By contrast, if the dark sector mass scale $m\sim{\cal O}(\rm MeV)$, we will get the correct order easily. But it immediately raises two questions: Is there any theoretical motivation for that light DM scale? And for that light DM how does it get correct relic density? Bare in mind that in the SM maybe only the photon and neutrino can be the final states of MeV scale DM annihilating, thus the second question concerns us.

For our FImP with SI, the two questions are simultaneously addressed in a coherent way. In the light of FImP miracle, interactions from Eq.~(\ref{Laga}) are supposed to freeze-in dark matter. Concretely, it is the two-body decay $\Phi\ra X X$ that dominates the freeze-in processes. And then the final yield can be formulated as~\cite{freezein1,Kang:2010ha}
\begin{align}\label{freezein}
Y_X(\infty)\approx{45\,g_\Phi\over1.66 \pi^4g_{*}^S \sqrt{g_{*}^\rho}}\f{\Gamma(7/2)\Gamma(5/2)}{16}\f{M_{\rm Pl}}{m_{\Phi}^2}\Gamma(\Phi\ra X X),
\end{align}
with $g_\Phi$ and $m_\Phi$ the internal degrees of freedom and mass of $\Phi$, respectively. The parameters $g_{*}^{S}$ ($g_{*}^{\rho}$ ) are the effective number of degrees of freedom contributing to the entropy (energy) density at $T\simeq m_\Phi$. Within the SM $g_{*}^{\rho}\approx g_{*}^{\rho}\approx106$. For multi mother particles contributing to freeze-in $X$, there is a summation over $\Phi$. Eventually, with $Y_X(\infty)$ one can express the FImP relic density as
\begin{align}\label{}
\Omega_{X} h^2=2.82\times10^5\L\f{m_{\rm DM}}{\rm MeV}\R Y_{X}(\infty).
\end{align}
In the ideal FImP miracle case, it is proportional to $\ld_\phi^{5/2}$ and $y_{\phi}^3$, for a real scalar and fermionic FImP respectively. For the typical scale of $u$ and $m_\Phi$, which are not far from the TeV scale, $\Omega_{X} h^2\simeq0.1$ uniquely determines the feeble couplings $\ld_{\phi}$ ($y_\phi$). In the following two subsections, we will detail the scalar and fermionic FImP with large self-interactions.

\subsection{The scalar FImP}

A scalar FImP $X=S$ can be easily realized in the scale invariant SM where only the Higgs doublet $H$ radiatively obtains VEV~\footnote{It is well known that this model fails in triggering successful electroweak spontaneously breaking and then modifications are indispensable. But this is not of our concern here~\cite{SI:boson,SI:hidden,SI:strong}. Our discussion is particularly suited for the modification where additional bosonic states are introduced to overcome the top quark.}. And the resulting model embodies the ideal FImP miracle. The relevant part of the model is very simple (see relevant studies of real scalar as a FIMP without SI~\cite{SI:S}):
\begin{align}\label{S:mass}
\f{\ld_{sh}}{2} S^2|H|^2+ \f{\ld}{4}S^4.
\end{align}
Note that here $S$ is automatically stable as a consequence of SI~\cite{Guo:2014bha}. This model leads to $u=v=246$ GeV. After EW spontaneously breaking (EWSB), the first term solely determines DM mass and relic density. The FImP mass is $m_S=\sqrt{\ld_{sh} /2}v$. Freezing-in of $S$ is dominantly proceeding via the decay of the SM Higgs boson $h$ into a pair of $S$, with decay width 
\begin{align}\label{}
\Gamma(h\ra SS)=\f{1}{32\pi}\f{\ld_{sh}^2 v^2}{m_h}.
\end{align}
Then, in terms of Eq.~(\ref{freezein}) the relic density is estimated to be
\begin{align}\label{S:relic}
\Omega_X h^2\simeq 0.12\times \L\f{\ld_{sh}}{10^{-10.5}}\R^{5/2} \L\f{v/m_h}{2.0}\R^{3}\L \f{10^3}{g_{*}^S \sqrt{g_{*}^\rho}}\R.
\end{align}
Equating $\Omega_X h^2$ with 0.11 one can fix the unique free parameter $\ld_{sh}$ and hence the mass of FImP, $m_S=1.0$ MeV. It was already obtained in the Appendix of our earlier work~\cite{Kang:2014cia}.

In most cases, the coupling constant $\ld$ plays no roles in DM phenomenologies. However, this ignored parameter is potential to generate a large DM self-interaction, especially for the FImP scenario under consideration. The resulting FImP self-scattering rate in per unit DM mass is given by
\begin{align}\label{}
\f{\sigma_{\rm DM}}{m_{S}}=\f{1}{128\pi}\f{\ld^2}{m_S^3}\simeq 7.9\times 10^{3}\L\f{\rm MeV}{m_S}\R^3\L\f{\ld}{0.1}\R^2\rm\,GeV^{-3}.
\end{align}
The scattering is from the contact interaction of $S$, so the scattering rate involves only the DM scale. Given a  light DM scale, the self-scattering rate easily becomes large as long as $\ld$ is not very small. This general advantage for light scalar dark matters, which always allow a quartic self-coupling to generate significant self-interactions, has already been utilized by the early references~\cite{Bento:2000ah}. We would like to stress again, not only the light DM scale but also correct DM relic density, which is fairly problematic for the MeV scale DM, are naturally and coherently achieved here by virtue of the FImP miracle.

\subsection{The self-interacting femionic FImP}

Although this example does not give an ideal FImP miracle, it takes a theoretical advantage, i.e., a Majorana FImP DM candidate $X=N$ is naturally predicted rather than introduced in the very low scale seesaw mechanism~\cite{RHN:FImP}. Its scale invariant version shows several theoretical merits~\cite{Kang:2014cia}. Scalar singlets $S_i$ with non-vanishing VEVs are necessary ingredients of the model, to generate Majorana mass for $N$. On the other hand, these singlets are also badly needed to implement hidden SI spontaneously breaking. At the same time, they alleviate the serious relic density problem of sterile neutrino DM through the freezing-in mechanism, admitting the FImP miracle (not ideal, see reasons later). Both are not far from the TeV scale.

Before heading toward the explanation to the not ideal miracle, here we brief the Lagrangian and report some relevant conclusions. We refer to Ref.~\cite{Kang:2014cia} for more details. Without imposing any symmetry by hand, the relevant Lagrangian takes a form of
\begin{align}\label{La:N}
-{\cal L}_N=V(S_i,H)+y_N\bar l HN+\f{\ld_i}{2}S_iN^2,
\end{align}
$V(S_i,H)$ is a generic scalar potential for the singlets and Higgs doublet, and its concrete form containing two singlets can be found in Ref.~\cite{Kang:2014cia}. In practice, at least two singlets (here we use the minimal number) are needed to trigger EWSB and moreover accommodate a quite SM-like Higgs boson near 125 GeV~\cite{2S:SI}. For later use, we denote the extra singlet-like Higgs bosons as ${H_2}$ and ${\cal P}$, with the latter (also the lighter) one being the pseudo Goldstone boson of SI spontaneously breaking.

The FImP miracle is not as ideal as that of the scalar FImP, mainly owing to the presence of multi singlets with VEVs. They provide multi sources for generations of FImP mass and relic,  whose numerical correlation hence is weakened. Even then, the miracle still holds in the sense of order of magnitude. Following Ref.~\cite{Kang:2014cia}, decays $H_a\ra NN$ freeze-in DM and its relic density can be parameterized as
\begin{align}\label{relic}
\Omega_{\rm DM} h^2=0.11\times\sum_{H_a={\cal P},H_2}\L\f{f_{H_a}^2}{1.0}\R
 \L\f{m_{\rm DM}}{0.1\,\rm MeV}\R^3\L\f{\rm 10\,TeV}{v_J} \R^2\L\f{1000\,\rm GeV}{m_{H_a}}\R\L \f{10^3}{g_{*}^S \sqrt{g_{*}^\rho}}\R,
\end{align}
where $m_{\rm DM}=M_{N}=\sum_i\ld_i \langle S_i\rangle$. Compared to Eq.~(\ref{S:relic}), the extra parameters $f_{H_a}$ (also the undetermined masses $m_{H_a}$) manifest the deviation from the ideal FImP miracle. They are model dependent, on the patters of singlets VEVs and as well their coupling to $N$. But in most cases they are order one numbers. Note that the fermionic FImP is favored to be in the much lighter region, around the sub-MeV scale or even below given that the singlets VEVs are not far above the TeV scale. The reason is nothing but that the fermion mass is proportional to the coupling constant $y_\phi$ instead of its square root like the scalar FImP. To maintain the coldness of the FImP DM, we take a conservative value $M_N=0.1$ MeV in this paper. Probably, it can be lowered down substantially, on account of a mildly colder DM spectrum from freeze-in~\cite{Merle:2015oja,Kang:2014cia}~\footnote{Despite beyond the scope of this paper, it is of interest to investigate the cosmological implications of warm DM with self-interaction. As shown here, a fermionic FImP with miracle tends to be a warm DM.}.

Now we turn our attention to self-interaction of $N$. Unlike the scalar FImP, large self-interactions are not a built-in part of the fermionic FImP. It calls for a light mediator, either a vector or scalar boson~\cite{Chu:2014lja}. Viewing from our Lagrangian Eq.~(\ref{La:N}), a light scalar boson, denoted as $S_0$, is a natural choice. $S_0$ is also a FImP, but it has a sizable coupling to $N$ via the Yukawa coupling ${\cal L}_{S_0}\supset -\ld S_0\bar NN$ (in four components). With it, $N$ can scatter with each other through $s/t/u-$channel exchanging $S_0$. To get a sufficiently large scattering rate, two options are of interest here. One is a heavy $S_0$ with mass $m_{S_0}$ much larger than $M_N$ and the other one is the opposite. In what follows we discuss them case by case.

\subsubsection{Dark force}

If $S_0$ is very light, it becomes a dark force mediator and DM self-scattering, in the non-relativistic limit, is described by the following attractive Yukawa potential
\begin{align}\label{}
V=-\f{\alpha_\ld}{r}e^{-m_{S_0}r}.
\end{align}
with $\alpha_\ld\equiv \ld^2/4\pi$. In different  parameter space spanned by $(\alpha_d, M_N, m_{S_0})$, the potential may induce different velocity-dependent DM self-scattering, and we refer to Ref.~\cite{Tulin:2013teo} for a comprehensive discussion. Here we focus on the simplest case, i.e., $\alpha_\ld M_N/m_{S_0}\ll1$ such that the Born approximation holds. Then, the perturbative computation in $\alpha_\ld$ from $V$ leads to~\cite{Tulin:2013teo}
\begin{align}\label{}
\sigma_T^{\rm Born}=\f{8\pi\alpha_\ld^2}{M_{N}^2v^4}\left[  \log\L1+\xi_v^2\R -\f{\xi_v^2}{1+\xi_v^2} \right],
\end{align}
with $\xi_v\equiv M_N v/m_{S_0}$. Note that in the small $M_N/m_{S_0}$ region such that $\xi_c\ll1$ one actually gets the velocity-independent approximation $\sigma_T^{\rm Born}\approx {4\pi\alpha_\ld^2M_{N}^2}/{m_{S_0}^4}$. If the self-interaction could leave observable effect at the cluster scale without spoiling the small scale structures, we should work in this limit~\footnote{Otherwise, self-scattering is over enhanced in the small scale system like dwarf galaxy with characteristic $v\sim10\rm\,km/s$ but is insufficient in the large scale system like cluster where $v\sim1000\rm\,km/s$.}. As an estimation, we typically need parameters as
\begin{align}\label{}
\f{\sigma_T^{\rm Born}}{M_N}=7.9\times10^3\L \f{M_{N}}{0.1\rm\,MeV} \R\L \f{0.002\rm\,MeV} {m_{S_0}}\R^4 \L\f{\alpha_\ld}{10^{-8}}\R^2\rm\,GeV^{-3}.
\end{align}
The resulted $\xi_v=50v$, which is indeed a small number for the typical velocity $v<10^{-3}$.

We make a comment on the fate of $S_0$. It is at the keV scale, assumed to gain mass as the scalar FImP in Eq.~(\ref{S:mass}). At leading order, it can decay into a pair of neutrinos, induced by the tiny active-sterile neutrino mixing, $\lesssim 10^{-10}$. The resulting decay lifetime $\gtrsim 10^{24}s$ is much longer than the cosmological timescale, so it survives as a relic today. But its energy fraction is negligible due to its lightness and small yield during freeze-in.

\subsubsection{Four-fermion interaction}

The other option is a heavy $S_0$, which actually leads to contact four-fermion interaction again, but here we make a direct calculation of the self-scattering without integrating out $S_0$. This process receives all $s/t/u-$channel contributions, and from them we get the following scattering rate per DM mass
\begin{align}\label{}
\f{\sigma_{\rm DM}}{M_{N}}=\f{3\ld^4}{8\pi}\f{M_{N}}{m_{S_0}^4}\simeq 6.0\times 10^{3}\L\f{M_N}{\rm 0.1\,MeV}\R\L\f{\rm MeV}{m_{S_0}}\R^4\L\f{\ld}{0.15}\R^4\rm\,GeV^{-3}.
\end{align}
$S_0$ is not favored to be much heavier than the MeV scale due to two reasons. One is that $\ld$ will become accordingly large, even larger than order 1,which is unpleasant at low energy. The other one is to prevent it from spoiling the FImP miracle. $S_0$ here is like a frozen-in scalar considered in Ref.~\cite{Merle:2013wta}, because itself is a FIMP and moreover could produce $N$ via decay $S_0\ra NN$. Therefore, the heavier $S_0$ means the larger yield of $S_0$ thus larger contribution to $N$ production. Let us estimate this from Eq.~(\ref{S:relic}) which just parameterizes the relic density of FImP $S$ with mass 1 MeV. As a naive estimation, one can get the $N$ relic density inherited from $S_0$ decay by multiplying Eq.~(\ref{S:relic}) a factor $M_N/m_{S_0}\sim0.1$. Thus, this contribution to the final relic density of $N$ is subdominant and the FImP miracle is not significantly affected. However, as we take heavier and heavier $S_0$, it will quickly 
dominate over the direct freeze-in production of $N$~\footnote{If $S_0$ gets mass by coupling to the singlet with VEV $v_s$ instead of Higgs doublet, in Eq.~(\ref{S:relic}) one should make the replacement $v\ra v_s$, and one will get a similar conclusion.}.

To end this paper we make a summarize. FImP is a necessary result after the combination between FIMP and SI, which further creates a FImP miracle. We show that a large DM self-interaction can be easily accommodated for FImP due to its lightness. This is particularly true for a scalar FImP which always has a quartic self-coupling. While for a fermionic FImP one has to introduce a mediator which is another FImP, a scalar boson with mass either much lighter or heavier than the FImP DM. DM self-scattering opens a new window to observe FImP (miracle), which does not leave traces in the conventional DM searches. For instance, they are potential to explain the recently observed DM self-interaction in the galaxy cluster Abel 3827.

In the late stage of this paper, we found that Ref.~\cite{freezein0} basically already studied the model of SM extended by a real scalar FImP, aiming at solving small scale problems using the self-interaction of FImP; Moreover, it pointed out that the MeV mass scale is consistent a model with zero bare Higgs mass, which is nothing but the classical scale invariance in our FImP framework.

\section*{Acknowledgment}
We would like to thank for helpful discussions with Yong Tang.




\end{document}